\input epsf
\documentstyle[prb,aps]{revtex}
\begin{document}
\draft
\title{RANDOM FIELD AND RANDOM ANISOTROPY EFFECTS IN DEFECT-FREE
THREE-DIMENSIONAL XY MODELS}
\author{Ronald Fisch}
\address{382 Willowbrook Drive\\
North Brunswick, NJ 08902}
\date{Submitted to Physical Review B, 23 November 1999}
\maketitle
\begin{abstract}
Monte Carlo simulations have been used to study a vortex-free XY
ferromagnet with a random field or a random anisotropy on simple cubic
lattices.  In the random field case, which can be related to a
charge-density wave pinned by random point defects, it is found that
long-range order is destroyed even for weak randomness.  In the random
anisotropy case, which can be related to a randomly pinned spin-density
wave, the long-range order is not destroyed and the correlation length is
finite.  In both cases there are many local minima of the free energy
separated by high entropy barriers.  Our results for the random field case
are consistent with the existence of a Bragg glass phase of the type
discussed by Emig, Bogner and Nattermann.

\end{abstract}
\section{INTRODUCTION}

The effects of random pinning on systems of charge-density waves (CDW) or
spin-density waves (SDW), and related problems like the pinning of the
Abrikosov vortex lattice, have been studied for a long time.\cite{RevCDW}
In real laboratory samples, there are always defects which create such
pinning forces.  Nevertheless, fundamental issues remain controversial.  In
1970, Larkin\cite{Lar70} presented an argument which shows that, if the
unpinned system is translation-invariant, (which means that we are ignoring
any effects of a periodic crystal-lattice potential), then weak random
pinning forces will destroy the long-range order (LRO) of a CDW in four or
fewer spatial dimensions.  A simpler domain wall energy argument was later
presented by Imry and Ma,\cite{IM75} and under some conditions it can be
made mathematically rigorous.\cite{AW90}  One unresolved issue is whether
these arguments can be extended to the SDW case, where there are
experiments which indicate the stability of LRO in the presence of
pinning.\cite{expts}  Another controversy involves the existence of the
proposed "Bragg glass" phase,\cite{GLD} which has quasi-long-range order
(QLRO).

If we ignore amplitude fluctuations,\cite{SP76,EL77} we can transform the
pinned CDW problem into an XY model in a random field, whose
Hamiltonian is usually taken to have the form
\begin{equation}
  H_{RFXY}= - J\sum_{<ij>}~\cos ( \theta_i - \theta_j )
            - G\sum_{i}~\cos (\theta_i - \phi_i )\, .
\end{equation}
Site $i$ is at position ${\bf r}_i$, the sites form a lattice, and $<ij>$
indicates a sum over nearest neighbors.  Each $\theta_i$ is a dynamical
variable representing the phase of the CDW at site $i$, and can take on
values in the interval $[ - \pi , \pi )$.  Each $\phi_i$ is a representation
of the random pinning energy arising from lattice defects.  Since the defect
sites are assumed to be immobile, the $\phi_i$ do not change with time.  We
also assume that the $\phi_i$ on different sites are uncorrelated, and that
the probability distribution for each $\phi_i$ is uniform on
$[ - \pi , \pi )$.

We can generalize Eq. (1) to study XY models in random $p$-fold fields,
where $p$ is any positive integer:
\begin{equation}
  H_{rp}= - J\sum_{<ij>}~\cos ( \theta_i - \theta_j )
            - G\sum_{i}~\cos ( p (\theta_i - \phi_i ))\, .
\end{equation}
For Eq. (2), each $\phi_i$ can be chosen to be in the interval
$[ - \pi / p , \pi / p )$, but $\theta_i$ still takes on values in
$[ - \pi , \pi )$.  The $p$ = 2 case, which is often simply referred to as
the XY model with random anisotropy, is related\cite{Fis95a} to a linearly
polarized SDW pinned by local moment impurities in the same way that the
$p$ = 1 case is related to a pinned CDW.  Note, in particular, that for
$p$ = 2 the Hamiltonian preserves a two-fold inversion symmetry, in
contrast to the $p$ = 1 case.  One consequence of this is that for $p$ = 2
(or more), unlike $p$ = 1, the time average of the local magnetization,
$\langle {\bf M} ({\bf r}_i , t ) \rangle_t =
\langle ( \cos ( \theta_i (t)), \sin ( \theta_i (t))) \rangle_t$,
must be zero for every $i$ in the paramagnetic phase.

Another model which is often considered is the elastic glass,\cite{VF84,GLD}
\begin{equation}
  H_{eg}= - J\sum_{<ij>}~( \theta_i - \theta_j )^2
            - G\sum_{i}~\cos ( p (\theta_i - \phi_i ))\, .
\end{equation}
For the elastic glass, the $\phi_i$ again have values in the interval
$[ - \pi / p , \pi / p )$, but the $\theta_i$ are now defined on
$( - \infty , \infty )$.  The dependence of Eq. (3) on $p$ is trivial, since
it can be removed by a rescaling of the variables.\cite{VF84}  Therefore,
after making the appropriate scaling, the behavior of the elastic glass
must be the same for all $p$.  Giamarchi and Le Doussal\cite{GLD} have
performed an analytical calculation which shows that in three dimensions
at zero temperature this model has a structure factor,\cite{Scomment}
\begin{equation}
  S_{\theta} ({\bf k}) = {1 \over L^3} | \sum_{j}^{L^3}~\theta_j
  \exp ( i {\bf k \cdot r}_j ) |^2 \, ,
\end{equation}
which diverges like $1 / |{\bf k}|^3$ at small $|{\bf k}|$.  This result has
recently been confirmed by a numerical simulation.\cite{MMZ99}

It is argued\cite{VF84,DFis97} that in the absence of topological defects
({\it i.e.} vortex lines), Eq. (1) should have the same continuum limit as
Eq. (3), and that this should be true for Eq. (2) as well.  The implication
is that the behavior of Eq. (2) should be essentially the same for all $p$,
just as is the case for Eq. (3).  However, the phase space for Eq. (3) is
simply connected, while that of Eq. (2) is not, even in the absence of
defects.  Since it is known that the numerical simulation results for the
$p$ = 3 case\cite{Fis92} of Eq. (2) do not show the same behavior as the
numerical simulation\cite{MMZ99} of Eq. (3), it appears that this difference
in the topology of the phase space for the two models invalidates the
mapping.  One can map the energies from Eq. (2) into Eq. (3), but the
entropies are different, even in the absence of vortex lines.

The difficulty is most transparent in the large $G / J$ limit considered by
Fisher.\cite{DFis97}  In this limit, for the elastic glass, Eq. (3),
each $\theta_i$ can still assume a countably infinite number of values, for
any $p$.  However, for the random field model, Eq. (2), each $\theta_i$ has
only $p$ distinct allowed values in this limit.  The $1 / |{\bf k}|^3$
divergence of $S_{\theta} ({\bf k})$ for the elastic glass arises from the
unbounded variations of the $\theta_i$.  This cannot occur in the random
field model, where $\theta_i$ is defined on a compact manifold.

\section{RANDOM $p$-FOLD FIELDS}

In the remainder of this work we will consider a version of the model of
Eq. (2), in the $p$ = 1 and $p$ = 2 cases.  In a ferromagnetic phase, where
time-reversal symmetry is spontaneously broken, the configuration averaged
value of ${\bf M} ({\bf r}, t)$ in a single sample,
$\langle {\bf M} ({\bf r}, t) \rangle_{\bf r}$, is not zero.  Thus,
naively,\cite{PPR78,PPRcomment} one would expect that if LRO is destroyed by
a weak random $p$ = 1 field, then it would also be destroyed by a weak
random $p$-fold field for any $p$.  This argument can be made explicit
within a perturbation theory\cite{PPR78} for small $G / J$ which is exact to
leading order in $1 / N$, where $N$ is the number of spin components.

The cause of the Larkin-Imry-Ma instability in the random field model may be
seen by studying the magnetic structure factor, whose form in three
dimensions is
\begin{equation}
  S ({\bf k}, t ) = {1 \over L^3} | \sum_{j}^{L^3}~{\bf M} ({\bf r}_j , t )
  \exp ( i {\bf k \cdot r}_j ) |^2 \, .
\end{equation}
In equilibrium, $S ({\bf k}, t)$ becomes independent of the time $t$ when
$L$ becomes infinite.  If the system is not ergodic, however, there may be
multiple equilibrium states, each with a different $S ({\bf k})$.  In a
ferromagnetic phase $S ({\bf k})$ shows a $\delta$-function peak at $\bf k$
= 0.  For the random $p$ = 1 field one shows\cite{IM75} that the presence of
this $\delta$-function induces a $1 / |{\bf k}|^4$ peak in $S ({\bf k})$ at
small $|{\bf k}|$.  Such a peak is impossible in four spatial dimensions or
less.  Due to the norm-preserving property of the Fourier transform,
\begin{equation}
  \sum_{\bf k}~S ({\bf k}) = L^3 \, ,
\end{equation}
where the sum over {\bf k} runs over the Brillouin Zone.  Since the square
of the length of each spin is one in this model, Eq. (6) merely states that
the total cross-section in a scattering experiment is equal to the number of
spins in the scattering volume times the cross-section of one
spin.  There is no corresponding sum rule for $S_{\theta} ({\bf k})$.

We proceed by separating the time-dependent and time-independent parts of
{\bf M}.  Without loss of generality, we can rewrite Eq. (5) in the form
\begin{equation}
  S ({\bf k}, t ) = {1 \over L^3} | \sum_{j}^{L^3}~(\langle
  {\bf M} ({\bf r}_j , t) \rangle_t + \delta {\bf M} ({\bf r}_j , t))
  \exp ( i {\bf k \cdot r}_j ) |^2 \, ,
\end{equation}
where $\langle \delta {\bf M} ({\bf r}_j , t) \rangle_t = 0$.  Performing
the Fourier transform and taking a time average yields
\begin{equation}
  \langle S ({\bf k}, t ) \rangle_t =
  |\langle {\bf M} ({\bf k} , t) \rangle_t |^2 +
  \langle |\delta {\bf M} ({\bf k} , t)|^2 \rangle_t \, .
\end{equation}
If the system is not ergodic, then we will find a different
$\langle S ({\bf k}, t ) \rangle_t$ for each equilibrium state.

To evaluate Eq. (8), Imry and Ma\cite{IM75} ignore the fixed-length-spin
constraint, and assume that for small $G / J$ they can use a linear response
spin-wave perturbation theory.  The second term of Eq. (8) is the standard
contribution from dynamical fluctuations of the spins.  In linear response
theory it gives a contribution to $S ({\bf k})$ of Lorentzian form,
proportional to $1 / (|{\bf k}|^2 +  1 / \xi^2 )$, where $\xi$ is the
correlation length.  When $G$ = 0, then $\xi$ is infinite in the
ferromagnetic phase, and within perturbation theory this remains true for
small $G / J$.

In a ferromagnetic phase, the linear response spin-wave theory for the first
term of Eq. (8) generates a $1 / |{\bf k}|^4$ peak whose amplitude is
proportional to $G^2$ times the square of the order parameter,
$\langle \langle |{\bf M} ({\bf r}_i , t)|^2 \rangle_{\bf r} \rangle_t$.
Such a peak is impossible in four dimensions or less, due to the sum rule on
$S ({\bf k})$.  If the decoupling of the different Fourier modes assumed in
the spin-wave approximation were valid, this would indicate the instability
of ferromagnetism in four dimensions or less in the presence of the random
$p$-fold field.  As discussed by Fisher,\cite{DFis85} this decoupling is
adequate when the number of dimensions is large, but it breaks down in four
dimensions or less.

For $p$ = 1 the random fields cause each $\langle {\bf M} ({\bf r}_i , t)
\rangle_t$ to become nonzero even in the paramagnetic phase.  Thus, for $p$
= 1 the spin-wave theory result in the paramagnetic phase for the first
term of Eq. (8) is a "Lorentzian-squared" peak, of the form $G^2 /
(|{\bf k}|^2 +  1 / \xi^2 )^2$.  The sum rule on $S ({\bf k})$ then implies
that $\xi$ must be finite.  This Lorentzian-squared peak also occurs for the
random field Ising model,\cite{IM75,GAAHS} and is not related to the
existence of massless spin waves.

There is no Lorentzian-squared peak in the paramagnetic phase for $p$ = 2,
since in this case each $\langle {\bf M} ({\bf r}_i , t) \rangle_t$ = 0, so
that the first term of Eq. (8) then makes no contribution to $S ({\bf k})$.
Therefore, the sum rule on $S ({\bf k})$ cannot prevent $\xi$ from diverging
for $p$ = 2, and the existence of a QLRO phase in this case was proposed in
1980 by Aharony and Pytte.\cite{AP80}

The domain wall energy scaling argument given by Imry and Ma,\cite{IM75}
which is nonperturbative, compares the relative strengths of the exchange
energy term and the random pinning term as a function of length scale.  If
the effective value of the coupling $G / J$ scales to infinity at large
length scales, then we know that for $p$ = 1 the model cannot be
ferromagnetic.

An analogous argument does not suffice for $p~>$ 1, however, because even
for strong random anisotropy the mean-field theory\cite{DV80} has a
ferromagnetic phase.  The domain wall energy argument does not account for
the exact $p$-fold symmetry of the Hamiltonian which exists for $p~>$ 1.
For $p~>$ 1 one cannot show that the random term uniquely determines the
large-scale structure of the low energy states.  Thus the rigorous proof
which works for $p$ = 1 cannot be applied for larger $p$.\cite{Aiz}

Because the spin-wave argument assumes replica symmetry,\cite{IM75,PPR78}
its lack of rigor has long been recognized.  More recently, M\'ezard and
Young\cite{MY92} have shown explicitly that when one calculates beyond the
leading order in $1 / N$ for $p$ = 1, the replica symmetry is broken in the
ferromagnetic phase, and that, therefore, the randomness should cause $\xi$
to be finite.  Since the randomness destroys translation invariance, it is
not surprising that it should also cause the long-wavelength spin-waves to
become massive.  Presumably this replica-symmetry breaking will also occur
for $p~>$ 1.

Within spin-wave perturbation theory,\cite{PPR78,DFis85} the effect of a
random anisotropy on the ferromagnetic phase appears to be the same as the
effect of a random field.  It seems reasonable that one should be able to
study the properties of a single minimum of the free energy, and that, at
least for small $G / J$, the behavior of the system in this local minimum
should not depend on $p$.  Because the replica symmetry is broken\cite{MY92}
for finite $N$, however, we know that this can fail.

There are a number of results which support the existence of LRO for XY
models ({\it i.e. N} = 2) with random anisotropy in four dimensions or less.
The first are the experiments\cite{expts} on SDW alloys which appear to have
LRO.  The second is the high temperature susceptibility series for the
random anisotropy XY model,\cite{FH90} which gives no indication of an
instability of ferromagnetism in four dimensions.  The third is the computer
simulations\cite{Fis92} for the $p$ = 3 case in three dimensions, which show
that the $p$ = 3 random anisotropy does not destroy the transition to
ferromagnetism, but the transverse correlation length in the ferromagnetic
phase becomes finite.

In this work we present the results of a computer simulation study of a
toy model which we believe preserves the essential features of Eq. (2).
For this model we find that in three dimensions a $p$ = 1 random field
perturbs the structure factor of finite lattices in a manner consistent
with the destruction of ferromagnetism for any strength of the random
field, as predicted by the domain wall energy scaling argument.  In the
corresponding $p$ = 2 case, however, we find no evidence for the destruction
of ferromagnetism.  Instead, we find that for this model the $p$ = 2 random
anisotropy causes $\xi$ to become finite without destroying the LRO.

\section{TOY MODEL FOR RANDOM FIELDS}

Large-scale computer simulations of the random field\cite{GH96,Fis97} and
random anisotropy\cite{Fis95a} XY models have been performed in the last few
years.  While the results of these simulations are quite instructive, it has
been difficult to study the behavior at weak randomness and low temperature.
This is due to the limited size of the lattices which can be studied, and
the difficulty of making transitions over energy barriers.  In order to
improve the effectiveness of the simulations, one may either try to develop
new techniques for studying Eq. (2), or else one may try to find a
modification of the Hamiltonian which preserves the essential features, but
is easier to study.  In this work we adopt the second approach.  We will
describe and study a model in which there are no energy barriers.

It was shown by Kohring, Shrock and Wills\cite{KSW86} that if one adds a
large vortex fugacity term to the XY model on a simple cubic lattice, then
the model retains a ferromagnetic equilibrium state even in the absence of
any explicit exchange energy.  It was later shown\cite{Fis95b} that this
"vortex-free" XY model, in which all allowed spin configurations have the
same energy, behaves in most respects like a normal XY model at some finite
temperature within the ferromagnetic phase.  Here, we will study the effects
of adding random fields and random anisotropies to the vortex-free model.
In order to retain the property that all allowed states of the model have
the same energy, we replace the random term of Eq. (2) by constraints on
each $\theta_i$.

To obtain a random-field-type model, for each $i$ we choose a random arc
of the circle of some fixed size, and declare that $\theta_i$ cannot take
on values within that arc.  The fraction of the circle which is removed
at each site is then a parameter which measures the strength of the random
field.  In order to maintain the vortex-free constraint everywhere, it is
sufficient that the fraction removed, $R_1$, be less than 1/2.  To see this,
note that any state in which all spins have values on the same half of the
circle, so that there is some axis for which the projection of all spins is
in the same direction, is vortex-free.  We will refer to such a state as a
"semicircle state".

For a random-anisotropy-type model we perform the same procedure, except
that we symmetrically remove two arcs from opposite sides of the circle
at each site.  In this case it is possible to satisfy the vortex-free
constraint even if at each site we only allow two points.  For the random
anisotropy model it is clear that the allowed states come in pairs, so
that time-reversal symmetry is not explicitly broken by the Hamiltonian.

We expect the qualitative behavior of the constraint-type random fields
and random anisotropies to be the same as the corresponding $p$ = 1 or
$p$ = 2 random terms of Eq. (2).  It is somewhat less clear that our
replacement of the exchange term in Eq. (2) by the vortex-free constraint
will not make any qualitative difference.  One can argue that for small
$G / J$ the low energy states of Eq. (2) should be vortex-free, but it is
difficult to prove this.  In the simulation of Gingras and Huse\cite{GH96}
it was observed that the vortex loops disappeared rapidly as the random
field was made weaker at low temperature.  It was suggested by them that in
the absence of vortex loops an XY model with a random field would possess a
QLRO phase, in which two-point correlations have a power-law decay as a
function of distance.  For the vortex-free model we can test this conjecture
in a straightforward manner.

\section{MONTE CARLO CALCULATION}

The Monte Carlo program was a modified version of one used
earlier\cite{Fis95b} to study the vortex-free model without randomness.  It
approximates the circle by a 256 state discretization, and uses a simple
cubic lattice with periodic boundary conditions.  Two linear congruential
pseudorandom number generators are used, one for assigning the random
fields, and a different one for flipping the spins.  The initial state of
each lattice is chosen to be a semicircle state.  Moves are rejected if they
would violate the vortex-free constraint or the local random-field
constraint.

A brief study of $L \times L \times L$ lattices as a function of size $L$
and the strength of the randomness showed that for the $p$ = 1 case
increasing the strength of the randomness caused a progressive decrease
of the equilibrium magnetization, $\langle \langle |{\bf M} ( L )|
\rangle_{\bf r} \rangle_t$, as expected, with $\langle \langle |{\bf M}|
\rangle_{\bf r} \rangle_t$ extrapolating to zero for large
$L$.  For the $p$ = 2 case, however, there was no evidence of a decrease of
$\langle |{\bf M}| \rangle_{\bf r}$ as the randomness was turned on.  In
order to investigate this unexpected result carefully, it was decided to
expend most of the computing effort on the computation of the structure
factor for lattices with $L$ = 64.

Starting from a semicircle state, each $R_1~>$ 0 lattice was run for 40,960
passes, which is several times the apparent longitudinal relaxation time.
Some of the $R_1$ = 0 lattices were run for only half this time, because the
longitudinal relaxation time is shorter in this case, and the transverse
relaxation is given by spin-wave theory.  The values of $\langle \langle
|{\bf M}| \rangle_{\bf r} \rangle_t$ were obtained by averaging over
the last half of each run, sampling every 20 passes.  For $L$ = 64, the
magnetization was found by this procedure to be
0.43516, 0.4018, 0.313 and 0.244 for $R_1$ = 0, 1/8, 1/4, and 3/8,
respectively.  The fluctuations in $\langle \langle |{\bf M}|
\rangle_{\bf r} \rangle_t$ between runs become larger as $R_1$ increases,
as does the time-averaged longitudinal susceptibility for a single run.
Because only one initial state was used for each $p$ = 1 sample, we do not
know if the variations in the time averages for different initial states of
the same sample are as large as the variations between samples.  The
transverse susceptibility , obtained from the time-dependent fluctuations of
$\langle {\bf M} \rangle_{\bf r}$ averaged over the last half of each run,
becomes smaller as $R_1$ increases.  For $R_1$ = 3/8 and $L$ = 64,
$\langle {\bf M} \rangle_{\bf r}$ remains close to its initial direction for
the duration of the run, and the transverse susceptibility is not much
larger than the longitudinal susceptibility.  This naturally implies that
there are many local minima of the free energy, at least on the time scale
of the simulation.  For $R_1$ = 1/8 and 1/4 the direction of
$\langle {\bf M} \rangle_{\bf r}$ may change substantially at first, but
then it seems to settle into some local minimum of the free energy, although
the transverse susceptibility remains large.

Results for the angle-averaged $S ( {\bf k} )$ for $L$ = 64 lattices with
these strengths $R_1$ of the random $p$ = 1 field are shown on a log-log
plot in Fig. 1.  Each data set is an average of eight samples of the
randomness, with one final state used for each sample.  The data for the
samples with no random fields approximately follow a $1 / |{\bf k}|^2$ law,
with an additional $\delta$-function at ${\bf k}$ = 0, which does not appear
on the log-log plot, as predicted by spin-wave theory.  As $R_1$ is
increased, the weight of the peak is progressively pushed out to larger
values of $|{\bf k}|$, with the sum rule on the integral over ${\bf k}$ of
$S ({\bf k})$ being preserved.  Because of the sum rule, the fluctuations in
the different small ${\bf k}$ modes are strongly coupled.  This makes it
difficult to estimate the statistical error for a single mode.  Suffice it
to say that the fluctuations of $S ({\bf k})$ for a single ${\bf k} \neq 0$
mode of a single sample are of about the same size as the average value for
that mode.

For $R_1$ = 1/8, the slope on the log-log plot of $S ( {\bf k} )$ in the
accessible small $|{\bf k}|$ region is approaching $-3$.  Due to the sum
rule, this indicates that there is no evidence for any LRO at this value of
$R_1$, even though the value of $\langle \langle |{\bf M}| \rangle_{\bf r}
\rangle_t$ is still not much reduced from its $R_1$ = 0 value at $L$ = 64.
At $R_1$ = 1/4, $S ( {\bf k} )$ shows an apparent slope of $-2.85 \pm 0.05$
for small $|{\bf k}|$ on the log-log plot.  By $R_1$ = 3/8 the $L$ = 64
samples are showing multiple local minima of the free energy.  This may be
an indication that some correlation length has become comparable to the
sample size, but $S ( {\bf k} )$ at $R_1$ = 3/8 can be fit at small
$|{\bf k}|$ by a power law in $|{\bf k}|$ with an exponent, $(-2 + \eta )$,
of $-2.63 \pm 0.07$.  A QLRO phase with a continuously varying value of
$\eta$ has recently been found in a similar model by Emig, Bogner and
Nattermann.\cite{EBN99}

In order to distinguish clearly between an infinite $\xi$ with a
continuously varying exponent $\eta$ and a finite $\xi$, we would need
data for larger $L$ or $R_1$ closer to 1/2. Either of these approaches
would require a substantial increase in computing effort.  If there really
is an infinite $\xi$ and an $\eta$ which varies continuously as a function
of $R_1$, then we would like to know if this behavior continues out to the
maximum allowed value of $R_1$, and, if so, how $\eta$ behaves near that
point.  If it were practical to perform simulations for larger values of
$L$, we believe that we would see the appearance of many local minima for
any nonzero allowed value of $R_1$.

To study the $p$ = 2 case, we concentrated on samples with $R_2$ = 3/8,
which means that only 1/4 of the states were allowed at each site.  Smaller
values of $R_2$ give an $S ({\bf k})$ almost indistinguishable from the
result for $R$ = 0, the model without randomness.  Note that $R_2$ = 3/8 can
be obtained from $R_1$ = 3/8 by removing an additional 3/8 of the allowed
states at each site, and thus restoring a two-fold symmetry.  Because the
random anisotropy constraints now cause most of the attempted moves to be
immediately rejected, each sample was run twice as long as for
$p$ = 1.  Also, two semicircle states, differing in average orientation by
$\pi / 2$ from each other, were used for each sample as initial states.

The average magnetization for $L$ = 64 and $R_2$ = 3/8 was obtained
by averaging over the last quarter of each run.  The result was
$\langle \langle |{\bf M}| \rangle_{\bf r} \rangle_t$ = 0.43613, slightly
${\it larger}$ than the result for the $L$ = 64 system with $R$ = 0.  The
results obtained using the two different initial states for a given sample
did not appear to be more similar to each other than results from two
different samples.  The direction of the magnetization rotated significantly
during most runs, indicating that $\xi$ is at least as large as $L$, and
that the observed behavior is unlikely to be due to a failure of the system
to relax.  For one of these eight samples the final states of the two runs
appeared to be in the same local minimum.  Studying smaller samples for much
longer running times also gave no indication that the results were caused by
insufficient relaxation.

Eight samples with maximal random anisotropy (only two allowed states at
each site, labeled $W_2$ = 1 in Fig. 2) were also studied.  In this case,
with only two allowed states per spin, a Metropolis-type algorithm, for
which spin flips allowed by the vortex-free constraint were made with
probability 3/4, was used to improve the efficiency of the program.  For
$W_2$ = 1, the $L$ = 64 value of $\langle \langle |{\bf M}| \rangle_{\bf r}
\rangle_t$ was 0.4662, and the orientation of $\langle {\bf M}
\rangle_{\bf r}$ always remained close to its initial direction.

The structure factor for these $p$ = 2 cases, again averaged over eight
samples, is shown in Fig. 2, along with the data for $R$ = 0.  We see that
at $L$ = 64 the structure factor for $R_2$ = 3/8 is not distinguishable from
that of $R$ = 0.  Although the data for $R_2$ = 3/8 appears to be slightly
above the data for $R$ = 0 at small $|{\bf k}|$, this is a sampling
artifact.  The actual average of $\langle |{\bf M}| \rangle_{\bf r}$ for the
16 $R_2$ = 3/8 states used in constructing the figure is 0.4349, while for
the 8 $R$ = 0 states it is 0.4357.

For $W_2$ = 1, $\xi$ is approximately 8 lattice spacings, and the lineshape
appears to be Lorentzian (plus the $\delta$-function at ${\bf k}$ = 0).  In
this case it appears that large $L$ behavior is seen already for $L$ = 16.
A sample with $W_2$ = 1 and $L$ = 16 was run for approximately $5 \times
10^5$ steps per spin.  The system appeared to relax to equilibrium within
the first 50 steps per spin, which was essentially the same as the
relaxation time for the $L$ = 64 samples, and no transitions were seen out
of the local minimum, which retained an $\langle {\bf M} \rangle_{\bf r}$
almost parallel to that of the initial semicircle state.  The transverse
susceptibility is approximately 15 times the longitudinal susceptibility.

\section{DISCUSSION}

The model we are using for our computer simulations is one in which all
allowed states of the spins have the same energy.  If the dynamical behavior
is ergodic, then, by definition, it must be possible to get from any initial
state to any final state.  One might imagine, therefore, that this model,
with single-spin flip dynamics, could have an ergodicity-breaking
transition as a function of the strength of the randomness.  That is, there
could be a transition between having a connected phase space to having a
phase space which is broken into many disconnected pieces.

It is easy to see, however, that any semicircle state can be connected to
another semicircle state with an arbitrary choice of the semicircle, by
single spin flips which do not violate the constraints.  This remains true
as long as $R_1$ or $R_2$ is less than 1/2.  Therefore, we cannot explain
the result that for large values of $R_1$ or $R_2$ we find many local
minima when $L$ becomes large by a percolation transition in phase space.
The breakdown of ergodicity is a true phase transition, because the
transition rates between different local minima only go to zero in the
infinite volume limit.

One should remember that the phase space available to an individual spin
depends on how well aligned its neighbors are.  In a semicircle state, the
magnetization is $2 / \pi$, about 0.63662.  Thus, when the system relaxes
into a state with $\langle |{\bf M}| \rangle_{\bf r}~<$ 0.5, the ability of
individual spins to reorient by single-spin flips is greatly reduced, even
though the entropy of the system as a whole has increased.

Entropy barriers\cite{Rit95} are just as effective as energy barriers in
suppressing transitions between different minima.  If the paths in phase
space between different local minima must pass through intermediate states
in which the value of $\langle |{\bf M}| \rangle_{\bf r}$ in a volume
$\xi^3$ is close to 0.6, then the probability of making such
transitions is suppressed by a factor exponential in the correlation volume.

The above estimate may be unduly pessimistic.  For example, it may be enough
to increase the local magnetization in a surface layer, so that the entropy
barrier is only proportional to $\xi^2$.  Nevertheless, the basic principle,
that uncorrelated single-spin flips are not an efficient way to achieve
large-scale reorientation of ${\bf M}$, is correct.

It would not be surprising if an alternative dynamics could be developed
which flipped large clusters of spins simultaneously,\cite{Wol89,Ros99} and
was thus more effective in moving through phase space.  Therefore, we would
like to check our results to see if they reflect true equilibrium behavior
by developing such an algorithm.  However, the results as they stand seem
internally consistent, and they are also consistent with the other related
results cited earlier.

For $p$ = 2, there is no instability of the LRO when the randomness is too
weak to induce the creation of vortex lines. We remind the reader that when
vortex lines are allowed, as for the strong random anisotropy limit of
Eq. (2), the LRO appears to be unstable in three dimensions, and the low
temperature phase seems to have only QLRO.\cite{Fis95a}  The nature of the
transitions between the LRO, QLRO and paramagnetic phases are clearly of
great interest, but they cannot be explored within the vortex-free model.

It should be noted that the infinite vortex fugacity used in our model does
not satisfy the smoothness conditions used in the proof of Aizenman and
Wehr.\cite{AW90}  Therefore, our finding that the random $p$ = 1 field
destroys the LRO in our model is an indication that the smoothness
conditions can be relaxed in three dimensions.  Of course, it does not
follow from this that the smoothness conditions can be relaxed in four
dimensions.

An alternative method\cite{PS92,Aiz94} of removing the vortices is by
placing a lower bound on the allowed values of
$\cos ( \theta_i - \theta_j )$ for all nearest neighbor pairs of $i$ and
$j$.  This method directly violates the smoothness condition (4.5) of
Aizenman and Wehr,\cite{AW90} and is a more severe constraint than the
vortex fugacity method used here.  It is likely that this alternative method
would produce results in qualitative agreement with those found here using
the Kohring-Shrock-Wills method.

\section{CONCLUSION}

In this work we have used Monte Carlo simulations to study a vortex-free
XY model in three dimensions with random $p$ = 1 and $p$ = 2 fields.  This
toy model is intended to represent the effects of random pinning on uniaxial
CDWs and SDWs.  We have found that for CDWs the LRO should be destabilized
by weak random pinning, but that for SDWs the LRO should survive.  These
conclusions are consistent with experiment.  Our results for the $p$ = 1
case are consistent with the existence of a QLRO of the type discussed by
Emig, Bogner and Nattermann.\cite{EBN99}

\acknowledgments

The author is greatful to Michael Aizenman and David Huse
for helpful discussions during the course of this work, and to the
physics departments of Princeton University and Washington University
for providing computing facilities.

\newpage
\begin{figure}
\caption{
Angle-averaged magnetic structure factor (per spin component) for the
vortex-free XY model with random fields on $64 \times 64 \times 64$ simple
cubic lattices, log-log plot.  Each data set shows an average of data from
8 samples.  The straight line has a slope of $-2$.}
\label{fig1}
\end{figure}

\begin{figure}
\caption{
Angle-averaged magnetic structure factor (per spin component) for the
vortex-free XY model with random anisotropy on $64 \times 64 \times 64$
simple cubic lattices, log-log plot.  Each data set shows an average of
data from 8 samples.  The $R_2$ = 3/8 data
set includes 2 states (with different initial conditions) per sample.
The straight line is identical to the one shown in Fig. 1.}
\label{fig2}
\end{figure}

\newpage
\begin{figure}
\epsfxsize=175pt 
\epsfbox{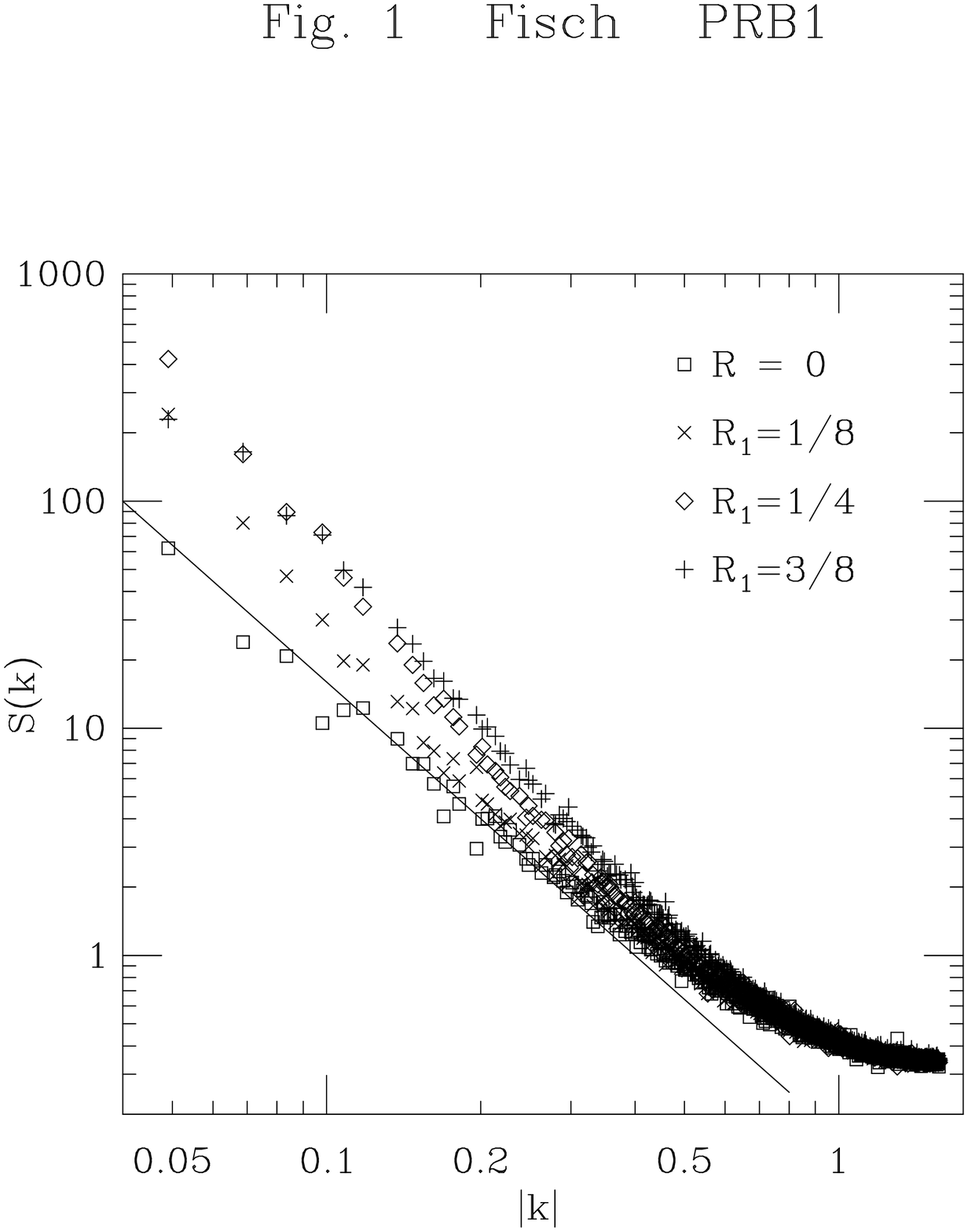}
\vskip 8truecm
\end{figure}

\begin{figure}
\epsfxsize=175pt 
\epsfbox{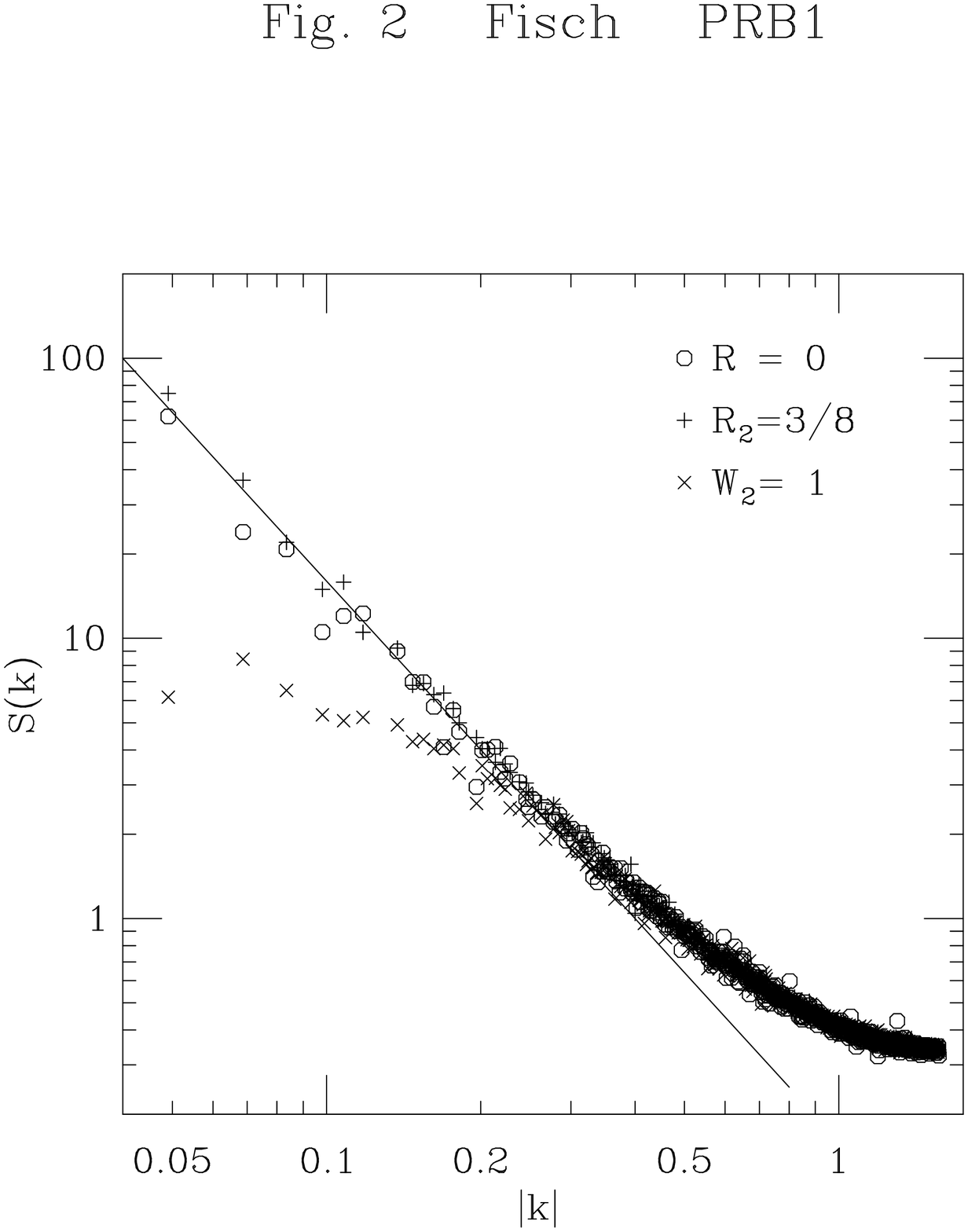}
\vskip 8truecm
\end{figure}


\begin{references}
\bibitem{Scomment}
The structure factor $S_{\theta} ({\bf k})$ which is calculated for Eq. (3)
is {\it not} the same as the magnetic structure factor $S ({\bf k})$ which
we will use for Eq. (2).  With Eq. (2) $S_{\theta} ({\bf k})$ is not gauge
invariant.
\end{references}
\end{document}